\documentclass[doublecol]{epl2}
\usepackage{graphicx,epstopdf}
\title{Amplitude death and resurgence of oscillation in network of mobile oscillators}
\shorttitle{Amplitude death in mobile oscillators} 

\author{Soumen Majhi and Dibakar Ghosh \footnote{e-mail: dibakar@isical.ac.in}}
\shortauthor{S. Majhi and D. Ghosh}

\institute{Physics and Applied Mathematics Unit, Indian Statistical Institute, Kolkata 700108, India\\
}
 {Complex systems }

\abstract{The phenomenon of amplitude death has been explored using a variety of different coupling strategies in the last two decades. In most of the work, the basic coupling arrangement is considered to be static over time, although many realistic systems exhibit significant
	changes in the interaction pattern as time varies.
	In this article, we study the emergence of amplitude death in a dynamical network composed of time-varying interaction amidst a collection of random walkers in a finite region of three dimensional space. We consider an oscillator for each walker and demonstrate that depending upon the network parameters and hence the interaction between them, global oscillation in the network gets suppressed. In this framework, vision range of each oscillator decides the number of oscillators with which it interacts. In addition, with the use of an appropriate feedback parameter in the coupling strategy, we articulate how the suppressed oscillation can be resurrected in the systems' parameter space. The phenomenon of amplitude death and the resurgence of oscillation is investigated taking limit cycle and chaotic oscillators for broad ranges of parameters, like interaction strength $k$ between the entities,  vision range $r$ and the speed of movement $v$. }

\begin{document}
\maketitle

 \section{Introduction}
Understanding a variety of emerging behaviors in networks of coupled oscillatory systems are very important due to its potential applicability in different fields of science including physics, biology, engineering and technology. The phenomenon of oscillation quenching \cite{ad} is one of the most important process among them. The interaction of oscillators in a complex network may result in a developmental phenomenon of complete annihilation of oscillations. It refers to the situation of stabilization of an already existing homogeneous steady state, which was otherwise unstable in the uncoupled systems.
 This unique emergent phenomenon is termed as amplitude death. Such behavior has enormous significance in many fields ranging from biological systems
 to laser physics. Amplitude death is particularly identified and characterized as a regulator to cease oscillation in neuronal systems, laser applications, electronic circuits etc. Various mechanisms leading to this oscillation quenching phenomena are elaborated in \cite{ad} and references therein.
 \par Most of the previous works on this phenomenon have assumed the interactions between the nodes to be invariant for all the course of time. However, due to its relevancy in diverse fields including physical, biological, social and engineering systems \cite{tempholm,timepnas,timepre}, especially in consensus problem \cite{cons}, power transmission system \cite{pts}, person-to-person communication \cite{mcom,mcom3,mcom2} etc, there have been strong urge in involving time-varying connections, particularly in dynamical networks \cite{belykh,sinha,sychen1,prasadth,prasadex,chaur,konish1,konish2,shen,movkur,movch,spect,infect,if1,pori,chaos2016} of late.
 \par We also note that there are many situations of movable physical entities in nature that can interact with each other depending strictly on their proximity. For example, a group of animals \cite{grani}, robotics \cite{robot}, motions in vehicles \cite{vehicle} correspond to such networks of particulars where mobility plays a crucial role. Again in chemotaxis \cite{chemo}, in wireless sensor networks \cite{sensor} and also in mobile ad hoc networks \cite{adhoc}, the process of mobility has enormous contribution. Although, there are works on investigation of synchronization in networks of movable oscillatory systems \cite{movkur,movch,spect,infect,if1,pori,chaos2016,ch2017}, but the influences of several physical system parameters related to mobile agents' network on the phenomenon of amplitude death are yet to be explored.
  \par As far as the fascinating process of oscillation quenching is concerned, in recent studies \cite{prasadth,prasadex,chaur,konish1,konish2}, the concept of amplitude death has been investigated theoretically  and experimentally in time-varying networks of static nodes. But studies so far, to the best knowledge of us, have mainly considered the variation of links over time, while keeping the spatial positions of the nodes fixed in space, except the work reported in \cite{shen} where mobility induced amplitude death in coupled oscillators is studied under metapopulation concept.
 \par In this article, we explore a new direction in amplitude death phenomenon arising in a network of mobile oscillators possessing random walk in a more accustomed way. We mainly focus on amplitude death phenomena in a mobile network where each node (agent) is randomly moving in a three dimensional (3D) physical space where the vision range of each agent decides the number of agents with which it interacts. We assume that each agent has an oscillator, periodic or chaotic. Depending on the network parameters, namely interaction strength $k$, vision range $r$ and velocity $v$ of the agents, amplitude death emerges. In this work, we presume a particular coupling form for all the course of time and there is no alteration in the coupling function. In this context, we would also like to mention that there are several processes, e.g., cardiac and respiratory systems \cite{new3}, El Ni\~{n}o/Southern Oscillation in Earth's ocean and atmosphere \cite{new2}, electric power generators \cite{new1}, where stable oscillations are required for proper functioning of the system.  Also there are so many real systems for which oscillation quenched states can be destructive and fatal. This suggests that there should be some proper prescription to resurrect oscillation from such quenched states. Regarding this issue of revival of oscillations, there exists some significant contributions \cite{zouprl,chaos17,pre17,res1,res2,res3,res4}. In the present article, we introduce a feedback parameter $\gamma$ in the coupling function for reviving the oscillation state from amplitude death state, the influence of which in doing so has been validated earlier \cite{res1,res2,res3,res4} for networks possessing stationary connections. 
  
  \par The rest of the paper is constructed as follows. In the next section, we discuss about the details of the random walk algorithm to explain how the oscillators move in three dimensional space. Then, we present our mathematical model for network of mobile oscillators where network topology changes due to oscillator's motion. After that, numerical results are shown for the mobile oscillators using limit cycle (Landau-Stuart) and chaotic (Lorenz) oscillators and show how oscillation gets suppressed depending upon the parameters like vision range, speed of movement and interaction strength. Finally, we summarize our results.

\section{Movement algorithm}
 We explain the scheme of random walk movement of oscillators in a finite arena of 3D space. Initially, $N$ number of  oscillators are randomly distributed in a $P\times Q\times R$ node mesh with a pseudo random number $R_i$ ( $0\le R_i\le1$) corresponding to the $i$-th  oscillator possessing position coordinate $(\xi_i,\eta_i,\zeta_i)$, $i=1, 2, \cdots , N$. 
 The movement of the oscillators in space is made through an elementary change in the position coordinates $(\xi_i,\eta_i,\zeta_i)$ after each time iteration in the following way:\\

(a) If the pseudo-random number $R_i$ corresponding to the $i$-th oscillator falls in  the interval $[0,\frac{1}{6})$ or $[\frac{1}{6},\frac{1}{3})$, then the movements along the +ve $x$-axis (the right) and -ve $x$-axis direction (the left) are made as :
\begin{equation}
\begin{array}{lcl}
\xi_i(\tilde t+\delta \tilde t)=\xi_i(\tilde t)+v~\mbox{cos}\theta $ [mod P]$,\\
\eta_i(\tilde t+\delta \tilde t)=\eta_i(\tilde t)+v~\mbox{sin}\theta $ [mod Q]$,\\
\zeta_i(\tilde t+\delta \tilde t)=\zeta_i(\tilde t)+v~ \mbox{sin}\theta $ [mod R]$,
\end{array}
\end{equation}
corresponding to $\theta=0$ or $\theta=\pi$ respectively, with $\delta \tilde t$  defining the characteristic time scale for the agents' motion. For simplicity, we assume that each agent moves in the space with the same speed $v$ during the entire course of time. Similarly, the movements along $y$ (to front or back) and $z$-axes (to up or down) corresponding to the cases of pseudo-random number $R_i$ lying in $[\frac{1}{3},\frac{1}{2})$ or $[\frac{1}{2},\frac{2}{3})$ and $[\frac{2}{3},\frac{5}{6})$ or $[\frac{5}{6},1]$ are defined, respectively.

\par (b) Updating time iteration, the described changes will be applied again for all the position coordinates $(\xi_i,\eta_i,\zeta_i)$ accordingly and so on.
 
\par (c) We note that for a particular instant, the oscillators interact with only those that belong to a specified region (subregion) of that particular oscillator depending on the proximity. So, inside the physical space of movement, we assign a much smaller region (subregion) of cubical structure to each and every oscillator. We call the subregion of an oscillator as vision size for each instant of time.
\par
\begin{figure}[ht]
\centerline{
\includegraphics[scale=0.55]{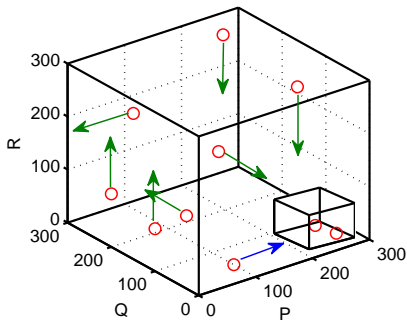}}
\caption{Moving oscillators (red circles) in three dimensional $P\times Q\times R$ node mesh. A cubical node mesh with $P=Q=R=300$ and $N=10$ for the clarity of the figure, has been chosen with arrows indicating their directions of movement. A subregion (vision size) of cubical structure is formed for a particular oscillator moving to its right, as directed by the blue arrow.}
\label{3dmove}
\end{figure}

 If at a particular instant of time, an oscillator is moving to its right (say) which is the case of the corresponding pseudo-random number to be in $[0,\frac{1}{6})$, then a cube of volume $r^3$ (vision size) will be created to its right. Then the oscillator will interact  only with those  oscillators which lie in the created cube. But it does not interact with any oscillator to its left at that instant of time. Actually, the vision size for a particular oscillator will be created in a direction along which the oscillator is moving. The same algorithm is applied to all the remaining cases. This type of creation of vision size is highly related with the situations of flashing fireflies, co-ordinated motions of vehicles etc. A simple graphical view of the movement of oscillators and creation of vision size at a particular instant of time is shown in Fig.~\ref{3dmove}.

\section{Mathematical form of the network}
\par We consider $N$ number of mobile agents moving in 3D space. Each agent $i~ (i=1,2,...,N)$ is associated with a dynamical system whose dynamics is defined by $\dot{X_i}=F(X_i)$, where $X_i$ is a $m$-dimensional vector of dynamical variables of the $i$-th oscillator and $F(X_i)$ is the corresponding vector field. We consider mathematical form of the network of mobile oscillators as:
\begin{equation}
\dot X_i=F(X_i)-k\bigg[ \gamma g_{ii}(t)(E_1 X_i)+\sum\limits_{j=1, j\not=i}^N g_{ij}(t)(E_2 X_j)\bigg ],
\end{equation}
where $k$ is the parameter signifying interaction strength between the moving nodes and $\gamma \;(0\le \gamma\le 1)$ is the feedback parameter that controls the diffusion. Here $G(t)=[g_{ij}(t)]_{N\times N}$ is the time-varying zero-row sum Laplacian matrix of order $N$ characterizing the network connections at the instant $t$.  Particularly, $g_{ij}(t)=-1$ if $j$-th oscillator falls in the vision range of $i$-th oscillator and zero otherwise, $g_{ii}(t)=\eta_i(t)$ where $\eta_i(t)$ is the number of agents lying in the vision size  of the $i$-th node at time $t$.    

 The matrices $E_1$ and $E_2$ are the $m\times m$ coupling matrices indicating which variables of one oscillator are coupled with which variables of the others.   Here $E_1$ and $E_2$ are chosen as (for the oscillators interacting with each other via dissimilar (conjugate) variables \cite{apconj} through the first two components),\\
$$ E_1 = \left( \begin{array}{ccccc}
  1 & 0 & 0 &\cdots & 0 \\
  0 & 1 & 0 & \cdots & 0 \\
  0 & 0 & 0 & \cdots & 0 \\
  \cdot & \cdot & \cdot & \cdots & \cdot\\
  0 & 0 & 0 & \cdots & 0
\end{array} \right)$$
 and
$$E_2 = \left( \begin{array}{ccccc}
  0 & 1 & 0 &\cdots & 0 \\
  1 & 0 & 0 & \cdots & 0 \\
  0 & 0 & 0 & \cdots & 0 \\
  \cdot & \cdot & \cdot & \cdots & \cdot\\
  0 & 0 & 0 & \cdots & 0
\end{array} \right).$$

\par In summary, the network is having three control parameters: the interaction strength between the agents $k$, the speed of movement of the agents $v$ and the vision (subregion) range $r$ of the oscillators. Without loss of generality, $P=Q=R=300$ with $N=100$ are taken throughout the work. In the rest of the paper, we will be concerned with how the network parameters $k, v$ and $r$ influence the system to realize global amplitude death and moreover we will also investigate how $\gamma$ affect the death state in broad ranges of those network parameters.
We have integrated the model (2) using fifth-order Runge-Kutta-Fehlberg algorithm with a time step size of $h=0.01$.   
\section{Results}
In this section, we investigate how the network model affect the dynamical behaviors of the moving limit cycle and chaotic oscillators where each of the oscillators are moving in 3D space. As a limit cycle oscillator, we take Landau-Stuart oscillator and as a chaotic oscillator we deal with paradigmatic  Lorenz oscillator.
 \par  Initially, at time $t=0,$ all the oscillators are randomly distributed in the physical space using the pseudo-random numbers $R_i$ $(i=1, 2, ... , N)$. 
 
We consider Landau-Stuart oscillators in the network (2) with $F(X_i)$ in the form\\
 \begin{equation}
 F(X_i)=\left(
 \begin{array}{c}
 (1-{p_i}^2)x_i-\omega y_i\\
 (1-{p_i}^2)y_i+\omega x_i\\
 \end{array}
 \right)
 \end{equation}
 where $p_i^2=x_i^2+y_i^2 ~~ (i=1, 2, ..., N)$ and the frequency $\omega=10$.
  In the uncoupled situation each Landau-Stuart oscillator exhibits a stable limit cycle near supercritical Hopf bifurcation and has an unstable focus at origin. The coupling matrices are $E_1=[1~~ 0; 0~~ 1]$  and $E_2=[0~~ 1; 1~~ 0]$.

\begin{figure}[ht]
\centerline{
\includegraphics[scale=0.4]{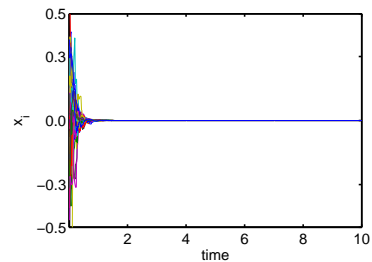}}
\caption{Time evolution of moving Landau-Stuart oscillators ($x$-components) for interaction strength $k=5$, speed of movement $v=30$ and vision range $r=50$.}
\label{adls}
\end{figure}

\par
\begin{figure}[ht]
	\centerline{
		\includegraphics[scale=0.42]{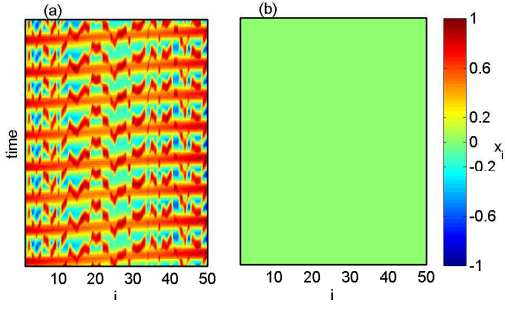}}
	\caption{ Spatio-temporal plots of moving Landau-Stuart oscillators: (a) incoherent state for $k=1.0$, (b) amplitude death state for $k=5.0$. Other parameters are same as in Fig.~\ref{adls}.}
	\label{lsst}
\end{figure}

 Figure~\ref{adls} shows the time evolution of all the moving Landau-Stuart oscillators for $v=30, r=50$ and $k=5$. From this figure we identify that the  oscillators when coupled, for certain values of parameters $v$, $r$ and $k$, cease to oscillate and get stabilized to the homogeneous steady state (i.e., origin) which was initially unstable. For further clarification, spatio-temporal plots are shown in Figs.~\ref{lsst} for different values of $k$ with fixed $v$ and $r$ that discriminate between the nature of oscillators before and after amplitude death. Figure~\ref{lsst}(a) shows incoherent (desynchronized) state of moving oscillators for $k=1.0$. For higher value of $k=5.0$, amplitude death state is shown in Fig.~\ref{lsst}(b).

\begin{figure}[ht]
	\centerline{
		\includegraphics[scale=0.65]{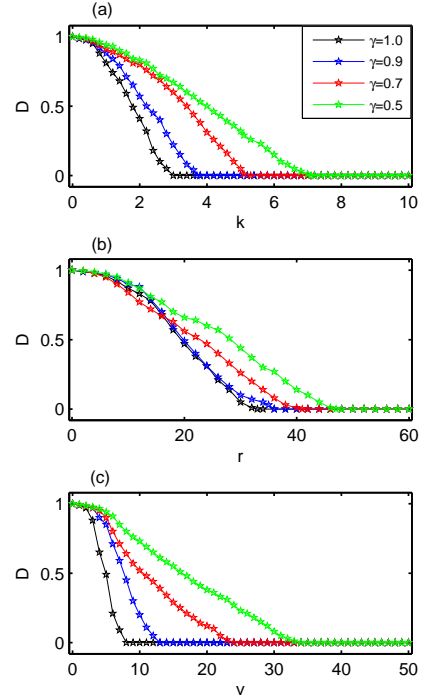}}
	\caption{(a) Normalized average amplitude $D$ with respect to the interaction strength $k$ with fixed $r=50$ and $v=20$ for different values of $\gamma=1.0, 0.9, 0.7, 0.5$ in black, blue, red and green curves respectively, as shown in inset. Normalized average amplitudes  against (b) vision range $r$ (for $k=10$, $v=20$) and (c) speed of movement $v$ (for $k=10$, $r=44$) for the above values of $\gamma$ are plotted. }
	\label{lsrv1d}
\end{figure}
To confirm death states, we calculate the average amplitude
of the network as the order parameter. The average amplitude $d$ of the coupled system is defined as\\
$d=\frac{1}{N}\sum\limits_{i=1}^N d_i ,~~\mbox{where}~ d_i=\langle x_{i,max} \rangle_t -\langle x_{i,min}\rangle_t~~ \mbox{for}~ i=1,2, ..., N $ and  $\langle\cdot \cdot \cdot \rangle_t$ represents average over time. The parameter $d$ becomes zero at the amplitude death state and non-zero values of $d$ signify oscillatory state. 
\par Now we plot the normalized value $D=\frac{d(k)}{d(0)}$ of the average amplitude as a function of the coupling strength $k$ for four different values of $\gamma$ in Fig.~\ref{lsrv1d}(a) with fixed $r=50$ and $v=20$. For $\gamma=1.0$, from the value of unity, $D$ (in color black) starts decreasing with increasing values of $k$ and becomes zero at $k=3.0$ resembling the state of global amplitude death and further remains zero for larger values of $k$. But, as smaller $\gamma=0.9$ is considered, $D$ (in blue color) remains non-zero unless $k\geq 3.7$. Thus smaller $\gamma$ helps the system to oscillate again even when the interaction strength $k$ passes the previous threshold value. For even smaller $\gamma=0.7$, $D$ (red color curve) continues to be non-zero and hence the oscillation persists until $k< 5.3$. For further decrement in $\gamma$ to $\gamma=0.5$, the oscillation sustains for any $k<7.1$. Thus $\gamma$ proves itself to be effective in resurrecting oscillation from death state whenever interaction strength $k$ varies.\\
 Next we move on to inspect the influence of $\gamma$ while the vision range $r$ varies with $k=10$ and $v=20$ fixed and plot $D=\frac{d(r)}{d(0)}$ as a function of $r$ in Fig.~\ref{lsrv1d}(b). Here again, for $\gamma=1.0$, $D$ decreases gradually with increasing $r$ and turns into zero at $r=33$, while for $\gamma=0.9$, $D$ becomes zero for a larger value of $r=36$. With $\gamma=0.7$, all the nodes continue oscillating even for $r=40$ and for $\gamma=0.5$, oscillation gets revived upto $r=46$.\\
 Finally, in Fig.~\ref{lsrv1d}(c), variations in $D=\frac{d(v)}{d(0)}$ with respect to the speed of movement $v$ for various $\gamma$ are shown. Keeping $k=10$ and $r=44$ unaltered, with $\gamma=1.0$, $D$ rapidly declines to zero at $v=8$. However, lowering $\gamma$ to $\gamma=0.9, 0.7, 0.5$, significant changes in the values of $v$ for which global amplitude death occurs, are observed. Those critical values of $v$ are $v=13, 24, 34$ respectively, implying high efficiency of $\gamma$ in order to restore oscillation from amplitude death.         
\begin{figure}[ht]
	\centerline{
		\includegraphics[scale=0.55]{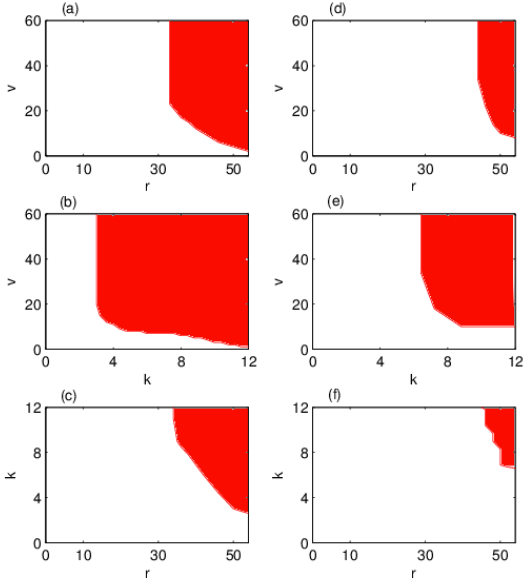}}
	\caption{Oscillation and amplitude death regions in the parameter space of (a,d) $v$ and $r$ with $k=10$ fixed, (b,e) $v$ and $k$ with $r=50$ fixed and (c,f) $k$ and $r$ with $v=20$, where for left panel (a,b,c), $\gamma=1.0$ and for right panel (d,e,f), $\gamma=0.5$. Red color region represents amplitude death while white sector corresponds to oscillatory state in all the figures. Here normalized average amplitude parameter $D$ is used to distinguish between oscillation and death states. }
	\label{lspsrv}
\end{figure}

 To examine the influence of all the network parameters, namely interaction strength $k$, speed of movement $v$ and vision range $r$ on cessation of oscillation in a broader prospect, we rigorously plot the parameter spaces for wide range of parameters in Fig.~\ref{lspsrv} that distinguishes between oscillation and amplitude death states.

Figure~\ref{lspsrv}(a) shows the $v-r$ parameter space plot for $N=100$ coupled LS oscillators with the interaction strength $k=10$ fixed for $\gamma=1.0$. Here the region with color red corresponds to amplitude death state while the arena with white color stands for oscillatory state. Here $D$ is used to distinguish between oscillatory and amplitude death states. As seen, for small values of $r$ and $v$, all the oscillators continue oscillating but for increasing $r\geq33$, amplitude death emerges and the critical value of $v$ needed for amplitude death decreases with growing $r$. The $v-k$ parameter space with $r=50$ is shown in Fig.~\ref{lspsrv}(b). Here again, initially for small values of $v$ and $k$, all the oscillators oscillate but as death starts appearing,  critical value of $v$ for the occurrence of amplitude death decreases for increasing $k$. Similar influence of the parameters are observed in Fig.~\ref{lspsrv}(c) where $k-r$ parameter space with $v=20$ is plotted.

\begin{figure}[ht]
	\centerline{
		\includegraphics[scale=0.38]{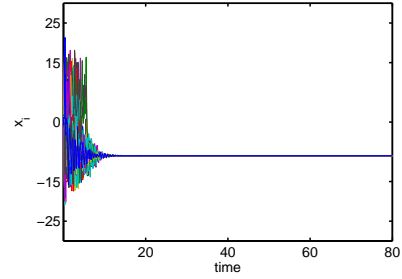}}
	\caption{Time evolution of mobile Lorenz oscillators ($x$-components) for interaction strength $k=10.0$, speed of movement $v=30$ and vision range $r=50$.}
	\label{adlor}
\end{figure}

Next we plot $v-r$ parameter plane for the LS oscillators' network at the same interaction strength $k=10$ fixed in Fig.~\ref{lspsrv}(d) with $\gamma=0.5$. Here the white region corresponds to oscillatory state while the region with color red represents amplitude death state, as before. As can be observed, for vision range $r<43$, amplitude death is not possible any more for any  speed of movement $v$ that effectively implies that the oscillation has been resurrected in that arena of the parameter plane. Figure~\ref{lspsrv}(e) shows $v-k$ parameter space keeping  $r=50$ fixed. Here again, the death island is observed to shrink considerably in the space. Finally, Fig.~\ref{lspsrv}(f) where $k-r$ parameter plane with $v=20$ is plotted, the amplitude death island gets reduced substantially in the parameter plane and only in a very small chunk of the space, death state persists.

 Finally, we consider moving Lorenz oscillators to see how moving chaotic units react under this type of interactional framework. We consider $N$ moving Lorenz oscillators in the form of equation (2) with 
 \begin{equation}
 F(X_i)=\left(
 \begin{array}{c}
 \sigma (y_i-x_i)\\
 x_i(\rho-z_i)-y_i\\
 x_i y_i-\beta z_i\\
 \end{array}
 \right) \\
 \end{equation}
  where $\sigma=10, \rho=28, \beta=\frac{8}{3}$ gives individual oscillator in chaotic regime.
We choose coupling matrices as $E_1=[1~~ 0~~ 0; 0~~ 1~~0;0~~0~~0]$  and $E_2=[0~~ 1~~0; 1~~ 0~~0;0~~0~~0]$. 

 \par We again discover that the chaotic oscillators coupled due to random movement in 3D physical space, for certain values of the parameters $v$, $r$ and $k$, cease to oscillate and get stabilized to one of the steady states $S_{\pm}=(\pm \sqrt{\beta (\rho-1)}, \pm \sqrt{\beta (\rho-1)}, \rho-1)$. Initially, at time $t=0,$ both of these steady states were unstable for each of the oscillators, for the above values of the system parameters but after their interaction, the stabilization of these steady states give rise to amplitude death. Figure~\ref{adlor} shows the time variation of all the $x-$components of moving chaotic agents for $v=30, r=50$ and $k=10.0$. It is noticed from this figure that all the oscillators get stabilized to the steady state $S_{-}=(-8.48, -8.48, 27)$.
 
 \begin{figure}[ht]
 	\centerline{
 		\includegraphics[scale=0.55]{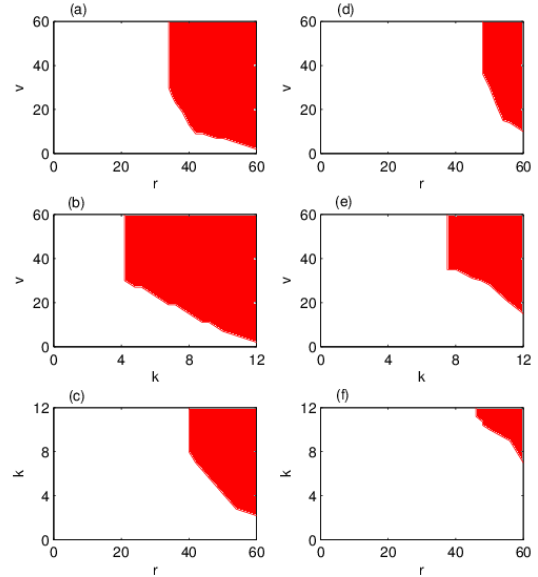}}
 	\caption{ Oscillation and amplitude death islands in the parameter space of (a,d) $v$ and $r$ with $k=10$ fixed, (b,e) $v$ and $k$ with $r=50$ fixed and (c,f) $k$ and $r$ with $v=30$, where for left panel (a,b,c), $\gamma=1.0$ and for right panel (d,e,f), $\gamma=0.5$ is taken. Red region resembles amplitude death island and white sector corresponds to oscillation in all the figures. For the distinction of death state from oscillatory state, normalized average amplitude $D$ is used.  }
 	\label{lorpsrv}
 \end{figure}

 Figure~\ref{lorpsrv}(a) depicts the $v-r$ parameter space for $N=100$ coupled  Lorenz systems with fixed interaction strength $k=10$. Here, as before, white and red regions respectively correspond to oscillation and  amplitude death states of the network. For small values of $r$ and $v$, all the oscillators continue to oscillate but for increasing $r\geq34$, amplitude death appears and the critical value of $v$ needed for amplitude death decreases for developing $r$. $v-k$ parameter space with $r=50$ is plotted in Fig.~\ref{lorpsrv}(b). Here again, initially for small values of $v$ and $k$, all the oscillators continue oscillating but as death starts appearing,  critical value of $v$ for the occurrence of amplitude death decreases for increasing $k$. Next in Fig.~\ref{lorpsrv}(c), $k-r$ parameter space with $v=20$ is shown and as seen, for $r\geq 40$, death can appear depending on $k$. 
 
 Figure~\ref{lorpsrv}(d) characterizes the $v-r$ parameter space for $N=100$ coupled  Lorenz systems with fixed interaction strength $k=10$ and for $\gamma=0.5$. For vision range $r<45$, amplitude death is not realized for any  speed of movement $v$ which suggests that the oscillation gets revived in that portion of the parameter plane. The $v-k$ parameter space keeping  $r=50$ fixed is shown in Fig.~\ref{lorpsrv}(e). Here again, the death island is observed to get shortened comprehensively in the space. Finally, Fig.~\ref{lorpsrv}(f) where $k-r$ parameter plane with $v=30$ is figured out, the amplitude death region gets reduced extensively and global amplitude death takes place only in a very small portion of the space.

\section{Conclusion}
To conclude, in this work we have considered a model of interaction among mobile oscillators moving in three dimensional space. One of the most fascinating emergent phenomenon in coupled oscillators' system, namely amplitude death in networks of spatially moving oscillators is realized depending upon the network parameters. We have examined amplitude death state in moving limit cycle and chaotic oscillators while most of the existing works are based on interactions that do not vary with time and even on the assumption of time evoloving communications in static nodes. But in many systems (e.g., communication, social, ecological networks etc.) nodes are not always static and  links are not always fixed, rather the connectivity between units may change during time due to mobility. Furthermore, we have successfully utilized the influence of a feedback parameter in the coupling form in order to revive oscillation from amplitude death state. We have identified the parameter regions for amplitude death and resurrected oscillatory states quite comprehensively by varying the three network parameters, namely the interaction strength $k$, speed of motion of the moving oscillators $v$ and vision range $r$. 
\begin{acknowledgments}
	D.G. was supported by the Department of Science and Technology of the Government of India (Grant EMR/2016/001039).
\end{acknowledgments}

\end{document}